\documentclass[%
reprint,
superscriptaddress,
 amsmath,amssymb,
 aps, authblk,
prb,
8pt]{revtex4-2}
\pdfoutput=1

\usepackage[pdftex]{graphicx}% Include figures
\usepackage{subcaption}
\usepackage{tikz}
\usepackage{adjustbox}
\usepackage{float}
\usepackage{dcolumn}% Align table columns on decimal point
\usepackage{physics}
\usepackage{siunitx}
\usepackage{bm}
\usepackage{bbold}
\usepackage{natbib}
\begin{document}

\preprint{APS/123-QED}

\title{Dirac Quantum Wells at Domain Walls in Antiferromagnetic Topological Insulators}

\author{N.B. Devlin}
\thanks{nbd22@cam.ac.uk}
\affiliation{Cavendish Laboratory, University of Cambridge, Cambridge CB3 0HE, United Kingdom}
\author{T. Ferrus}
\affiliation{Hitachi Cambridge Laboratory, Cambridge CB3 0HE, United Kingdom}
\author{C.H.W. Barnes}
\affiliation{Cavendish Laboratory, University of Cambridge, Cambridge CB3 0HE, United Kingdom}

\date{\today}

\begin{abstract}

We explore the emergence of spin-polarised flat-bands at head-to-head domain walls in a recently predicted class of antiferromagnetic topological insulators hosting planar magnetisation. We show, in the framework of quantum well physics, that by tuning the width of a domain wall one can control the functional form of the bound states appearing across it. Furthermore, we demonstrate the effect that the parity of the number of layers in a multilayer sample has on the electronic dispersion. In particular, the alignment of the magnetisation vectors on the terminating surfaces of odd layer samples affords particle-hole symmetry leading to the presence of linearly dispersing topologically non-trivial states around \(E = 0\). By contrast, the lack of particle-hole symmetry in even layer samples results in a gapped system, with spin-polarised flat-bands appearing either side of a band gap, with characteristic energy well within terahertz energy scales. In addition to being a versatile platform for the development of spintronic devices, when many-body interactions are accounted for we predict that these flat-bands will host strong correlations capable of driving the system into novel topological phases. 
    
\end{abstract}

\maketitle

\section{Introduction}

Topological Insulators (TIs) are bulk insulating materials possessing an inverted band structure that exhibits helical surface states with a linear dispersion centred at a Dirac point \cite{fu_topological_2007,xia_observation_2009,hasan_colloquium:_2010}. These surface states are afforded topological protection by time-reversal symmetry (TRS) and are insensitive to elastic backscattering by time reversal invariant perturbations, such as electric fields from impurities and dopants \cite{zhang_experimental_2009,roushan_topological_2009}. In the decade since their discovery TIs have been the subject of intense experimental and theoretical research and have been proposed for use in low-power electronics, rapid magnetisation switching \cite{fan_magnetization_2014,wang_room_2017,li_magnetization_2019} and quantum computation \cite{fu_superconducting_2008,lian_topological_2018}.
    
The interplay between topology and magnetism is known to break TRS, resulting in numerous exotic phases of matter \cite{essin_magnetoelectric_2009,yu_quantized_2010,chang_experimental_2013,wang_quantized_2015,he_chiral_2017,xiao_realization_2018}. Initial methods aimed at inducing ferromagnetism in the Dirac fermions included the introduction of Cr/V magnetic dopants, but owing to the large inhomogeneity in the distribution of these dopants across the TI, sub-Kelvin temperatures must be accessed to establish sufficient long-range magnetic order to realise these exotic states of matter \cite{mogi_magnetic_2015,li_origin_2016,ou_enhancing_2018,wu_stabilizing_2019}. In recent years a new class of intrinsically magnetic TIs have been proposed and realised; antiferromagnetic TIs (AFMTIs) \cite{otrokov_highly-ordered_2017,eremeev_new_2018,gong_experimental_2019,wu_natural_2019,otrokov_prediction_2019,qi_pursuing_2020}. These van der Waals layered materials are composed of septuple layers of the form \(\text{XB}_{2}\text{T}_{4}\), where X is a magnetic transition/rare earth element, \(\text{B} = \text{Sb, Bi}\) and \(\text{T} = \text{Se, Te}\). Within each layer spins order ferromagnetically, while the interlayer coupling is antiferromagnetic, that is, A-type antiferromagnetism. 

\begin{figure}[ht!]
    \centering
    \begin{subfigure}[b]{0.17\textwidth}
    \begin{tikzpicture}
        \node[anchor=south west,inner sep=0] (image) at (0,0) {\includegraphics[width=1.0\textwidth]{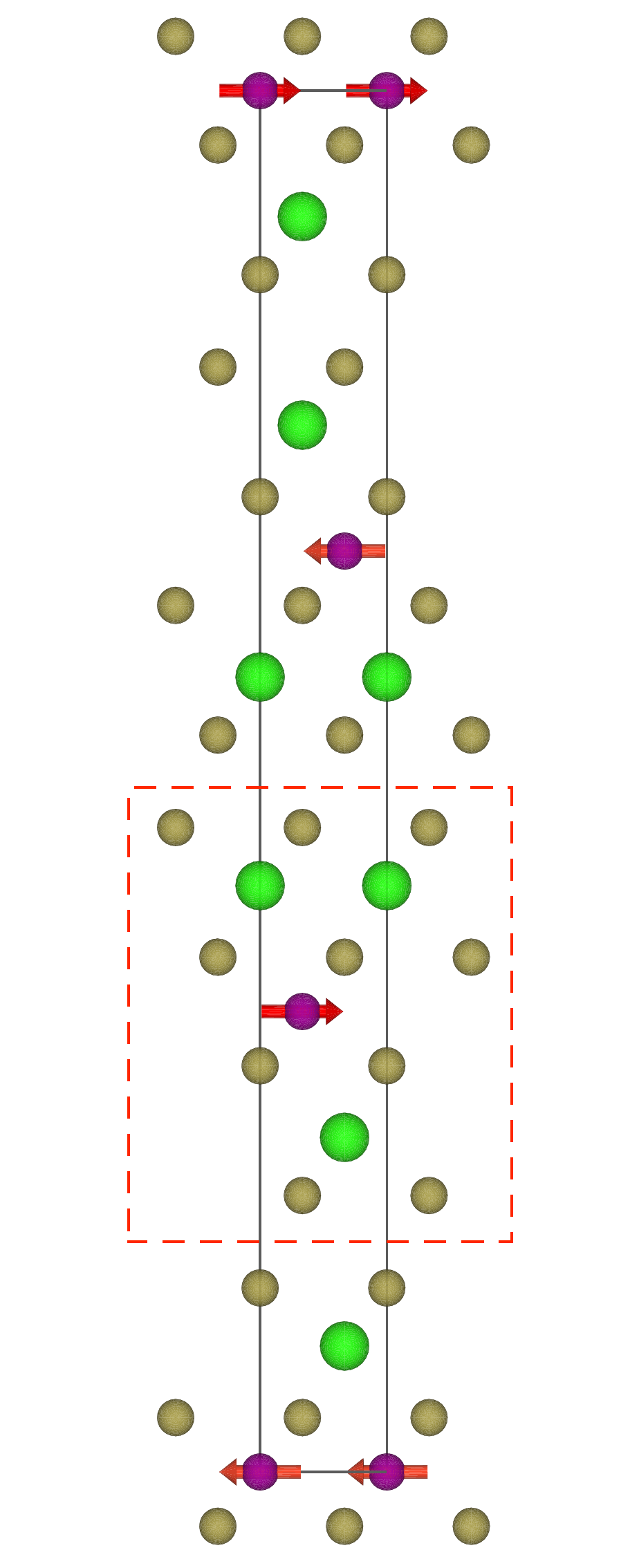}};
        \begin{scope}[x={(image.south east)},y={(image.north west)}]
            %\draw[help lines,xstep=.1,ystep=.1] (0,0) grid (1,1);
            %\foreach \x in {0,1,...,9} { \node [anchor=north] at %(\x/10,0) {0.\x}; }
            %\foreach \y in {0,1,...,9} { \node [anchor=east] at %(0,\y/10) {0.\y}; }
            \draw[red,dashed,rounded corners] (0.20,0.20) rectangle (0.80,0.50);
            \draw[black] (0.0,0.81) node {(a)};
        \end{scope}
    \end{tikzpicture}
    \end{subfigure}
    \begin{minipage}[b]{0.22\textwidth}
    \begin{subfigure}[b]{\linewidth}
    \begin{tikzpicture}
        \node[anchor=south west,inner sep=0] (image) at (0,0) {\includegraphics[width=1.0\textwidth]{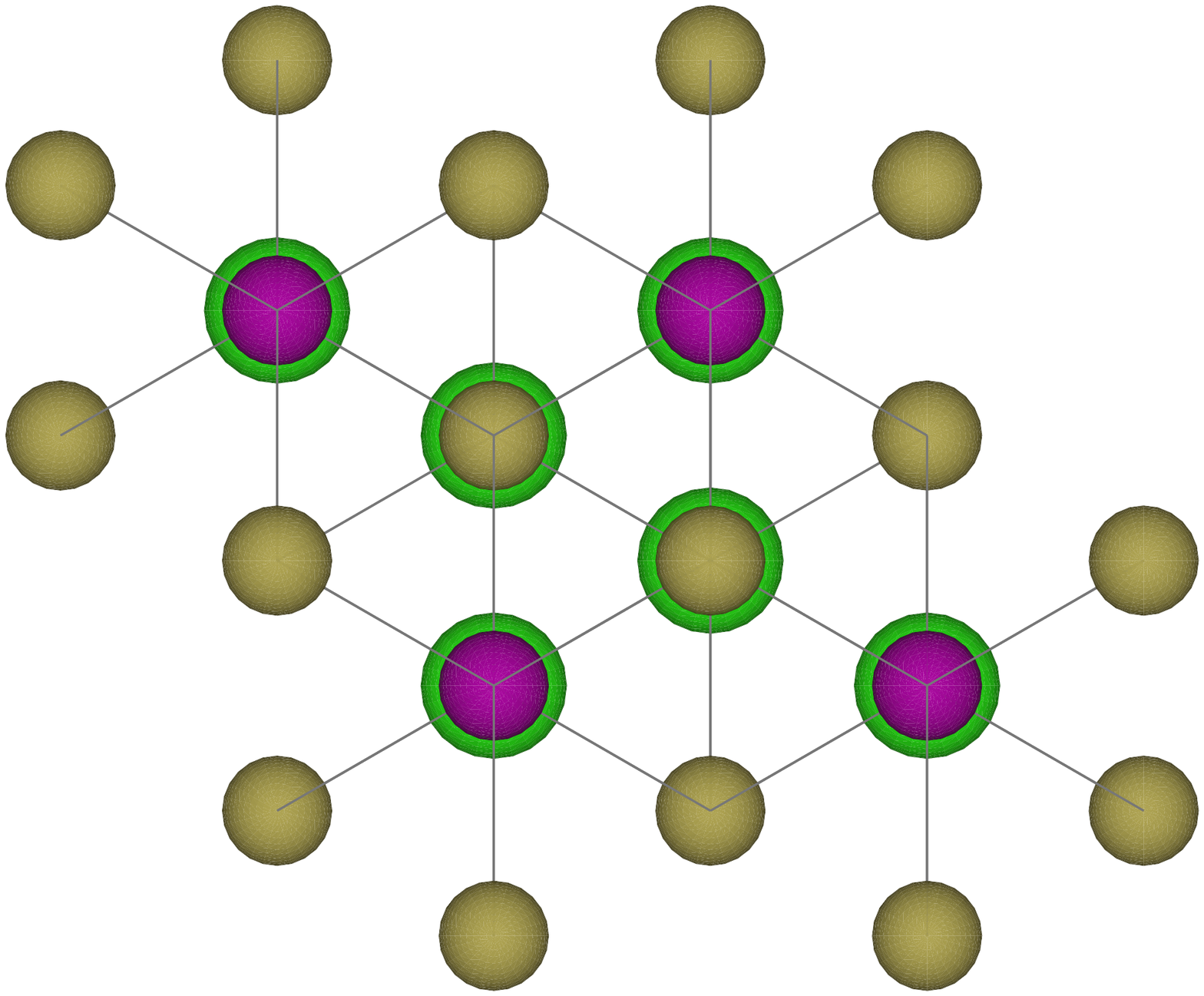}};
        \begin{scope}[x={(image.south east)},y={(image.north west)}]
            %\draw[help lines,xstep=.1,ystep=.1] (0,0) grid (1,1);
            %\foreach \x in {0,1,...,9} { \node [anchor=north] at %(\x/10,0) {0.\x}; }
            %\foreach \y in {0,1,...,9} { \node [anchor=east] at (0,\y/10) %{0.\y}; }
            \draw[black] (0.0,1.0) node {(b)};
        \end{scope}
    \end{tikzpicture}
    \end{subfigure}
    \begin{subfigure}[b]{\linewidth}
    \begin{tikzpicture}
        \node[anchor=south west,inner sep=0] (image) at (0,0) {\includegraphics[width=0.5\textwidth]{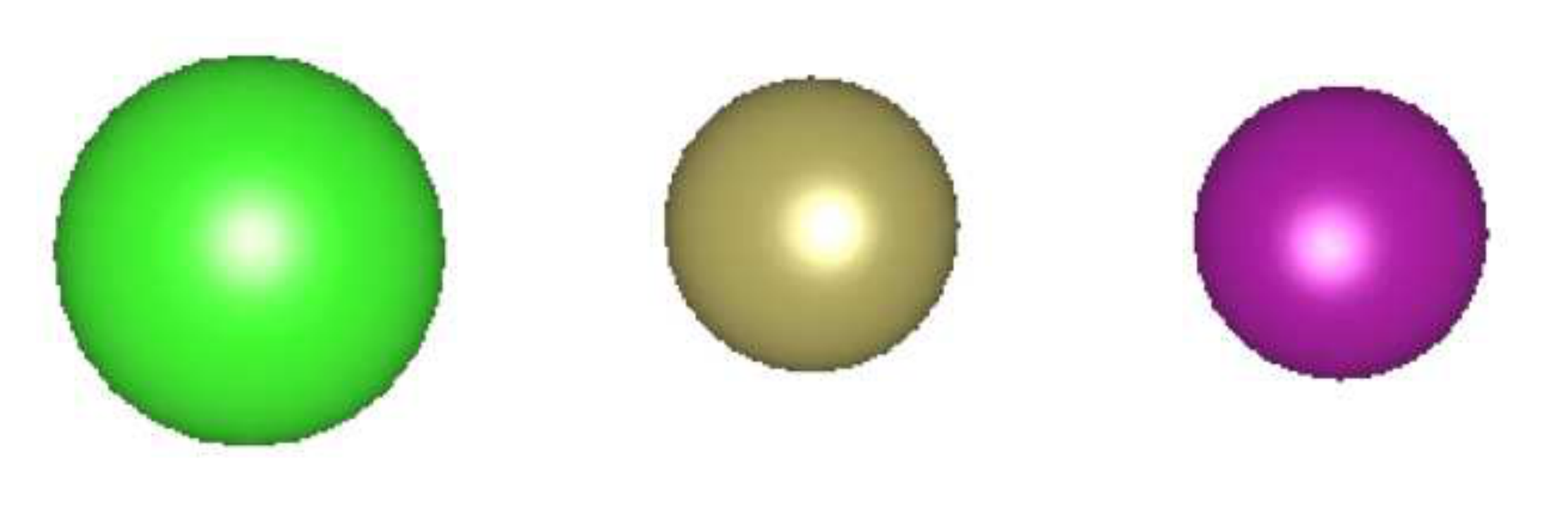}};
        \begin{scope}[x={(image.south east)},y={(image.north west)}]
            %\draw[help lines,xstep=.1,ystep=.1] (0,0) grid (1,1);
            %\foreach \x in {0,1,...,9} { \node [anchor=north] at (\x/10,0) {0.\x}; }
            %\foreach \y in {0,1,...,9} { \node [anchor=east] at (0,\y/10) {0.\y}; }
            \draw[black] (0.17,-0.15) node {\footnotesize Bi};
            \draw[black] (0.54,-0.15) node {\footnotesize Te};
            \draw[black] (0.90,-0.15) node {\footnotesize V/Eu};
        \end{scope}
    \end{tikzpicture}
    \end{subfigure}\\[\baselineskip]
    \end{minipage}
    \caption{Crystal structure of \(\text{VBi}_{2}\text{Te}_{4}\)/\(\text{EuBi}_{2}\text{Te}_{4}\) along the (a) \(a\)-axis and (b) \(c\)-axis. The crystal structure is rhombohedral with the space group \(R\bar{3}m\). The crystal is a layered structure formed of septuple layers (red box) separated by a van der Waals gap. The paramagnetic unit cell (black box) is shown in (a), however it should be noted that the A-type antiferromagnetism of this material leads to a magnetic unit cell twice the length of that shown. This figure was produced using VESTA \cite{momma_vesta_2011}.}
    \label{fig:VBi2Te4_crystal}
\end{figure}
    
While ferromagnetism breaks TRS, this A-type antiferromagnetism preserves the symmetry given by \(\tau_{\frac{1}{2}}\mathcal{T}\) where \(\mathcal{T}\) is the TRS operator and \(\tau_{\frac{1}{2}}\) represents a primitive lattice translation \cite{mong_antiferromagnetic_2010}. This additional symmetry permits the existence of topological phases in AFMTIs that would otherwise be impossible in ferromagnetic systems. Recent research has focused heavily on the compound \(\text{MnBi}_{2}\text{Te}_{4}\), which orders along the \(z\) axis (that is, along the growth axis), and is viewed as the most promising candidate for observation of a high-temperature quantum anomalous Hall effect (QAHE) \cite{yu_quantized_2010,chang_experimental_2013,chen_intrinsic_2019,deng_quantum_2020,zhang_topological_2019}. However, the family of \(\text{XB}_{2}\text{T}_{4}\) AFMTIs is extensive and other spin configurations are possible. In particular, in-plane magnetisation has been predicted in V and Eu based AFMTIs \cite{li_intrinsic_2019,petrov_domain_2020}, shown in Figure \ref{fig:VBi2Te4_crystal}. While out-of-plane magnetism opens a gap in the topological surface states, in-plane magnetism preserves the gapless nature of these states but shifts the position of the Dirac point in momentum space by an amount proportional to the magnetisation.

\begin{figure}
    \centering
    \begin{tikzpicture}
        \foreach \x in{0,...,4}
    {   \draw[black,ultra thick] (0,\x,4) -- (4,\x,4);
        \ifnum\x=0
        \draw[black, ultra thick] (\x,0,4) -- (\x,4,4);
        \fi
        \draw[black, ultra thick] (4,\x,4) -- (4,\x,0);
        \ifnum\x=0
        \draw[black, ultra thick] (\x,4,4) -- (\x,4,0);
        \fi
    }
    %\draw[black, ultra thick] (4,0,0) -- (4,4,0);
    \draw[black, ultra thick] (4,0,4) -- (4,4,4);
    %\draw[black, ultra thick] (0,4,0) -- (4,4,0);
    \draw[black, dashed] (2,0,4) -- (2,4,4);
    \draw[black, dashed] (2,4,0) -- (2,4,4);
    
    \foreach \y in{0.5,2.5}
    {   \draw[-latex, ultra thick] (0.5,\y,4) -- (1.5,\y,4);
        \draw[latex-, ultra thick] (2.5,\y,4) -- (3.5,\y,4);
    }
    \foreach \y in{1.5,3.5}
    {   \draw[latex-, ultra thick] (0.5,\y,4) -- (1.5,\y,4);
        \draw[-latex, ultra thick] (2.5,\y,4) -- (3.5,\y,4);
    }
    
    \draw[-latex] (-2.0,-0.0,4.0) -- (-2.0,-0.0,3.0);
    \draw[black] (-2.0,-0.0,2.7) node {\(x\)};
    \draw[-latex] (-2.0,-0.0,4.0) -- (-1.0,-0.0,4.0);
    \draw[black] (-0.8,-0.0,4.0) node {\(y\)};
    \draw[-latex] (-2.0,-0.0,4.0) -- (-2.0,1.0,4.0);
    \draw[black] (-2.0,1.3,4.5) node {\(z\)};
    
    \end{tikzpicture}
    \caption{Schematic of a multilayer \(\text{VBi}_{2}\text{Te}_{4}\)/\(\text{EuBi}_{2}\text{Te}_{4}\) sample with a head-to-head (tail-to-tail) domain wall. In this article, we will consider samples of infinite length along the \(x\) axis. The magnetisation in each septuple layer either side of the domain wall (dashed line) is shown by black arrows. A sharp domain wall is shown, however it should be noted that competition between the magnetocrystalline and exchange energies will lead to a realistic domain wall having a finite width.}
    \label{fig:head-to-head_domain}
\end{figure}
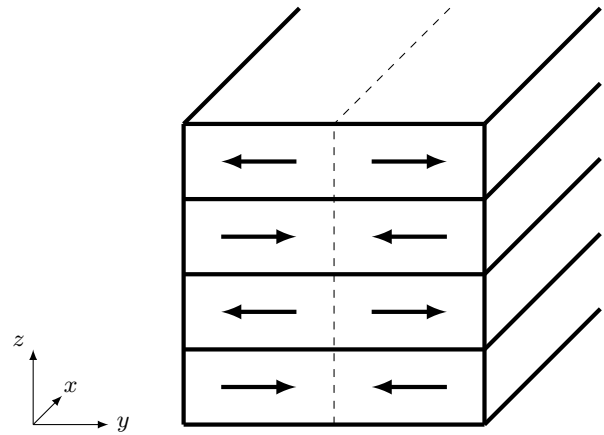

The formation of domain walls (DWs) in magnetic samples can be attributed to the minimisation of competing energies, namely the magnetocrystalline anisotropy and the exchange energy. DWs have a variety of applications in the development of disruptive device technologies. Furthermore, the ability to rapidly reconfigure DWs in magnetic samples by non-destructive means makes them an attractive playground to investigate exotic spin phenomena \cite{yasuda_quantized_2017,rosen_chiral_2017}. Indeed, recent work has suggested the existence of quasiparticles with infinite effective mass at head-to-head (equivalently, tail-to-tail) DWs in AFMTIs with in-plane magnetisation \cite{petrov_domain_2020}. A schematic of head-to-head (equivalently, tail-to-tail) DWs in a multilayer sample is shown in Figure \ref{fig:head-to-head_domain}. The existence of these quasiparticles may be inferred through a topological argument. For in-plane magnetisation, the Dirac cones will be shifted by equal and opposite amounts in momentum space either side of a DW. The continuity of the energy dispersion relation demands that there must be a band of bound states joining these two Dirac cones across the DW. Furthermore, in order to respect the spin-momentum locked nature of the Dirac fermions, such electronic bands must be spin-polarised flat-bands, i.e. bands with an infinite effective mass. Throughout this work, we focus on the case of transverse head-to-head (tail-to-tail) DWs. While such DWs are more experimentally challenging to realise and control than vortex DWs \cite{vanhaverbeke_control_2008}, they offer the advantage of a non-zero in-plane stray field which is a necessary requirement for DW detection in any device \cite{parkin_magnetic_2008}.

It is important to note that discussions involving symmetry protected topological phases are typically based on a single-particle picture, where interactions are neglected.  However when many-body interactions are accounted for, flat-bands, which we argued must exist across a transverse head-to-head DW, provide an ideal platform to investigate the emergence of collective phenomena due to strong correlations. While the scope of this article is limited to a single-particle description, we note that the interplay between topology, magnetism and strong electron-electron interactions (EEI) could give rise to even more exotic states of matter. Tuning the Fermi level to alter the filling of low energy flat bands will alter the competition between these effects, giving rise to a rich topological phase diagram. Furthermore, just as the parity of the number of layers in AFMTIs with spins aligned along the \(z\) direction determines whether the sample is in the QAHE or Axion insulator states \cite{otrokov_unique_2019}, we should expect an analogous effect in samples with in-plane magnetism.
    
The structure of this paper is as follows: in section II we  explicitly show the formation of interfacial states centred at transverse head-to-head DWs, by working within the framework of quantum well (QW) physics. In particular, by varying the smoothness of the DW we find that the functional form of the QW and the emergent bound states can be controlled. Because of their technological importance within the area of device physics, we focus on samples of finite size. In section III we construct a model multilayer AFMTI Hamiltonian. Using this model, we show that the electronic dispersion is dependent on the parity of the number of layers. By exploiting the structural inversion symmetry of samples with odd numbers of layers we show that low energy edge-states protected by particle-hole symmetry are present in these samples but not in even layered samples, where there is a material dependent band gap between spin polarised flat-bands.

\section{Surface Hamiltonian}

Before we examine the more complex situation of an AFMTI, we will examine the existence of DWs within a magnetic TI using a simple  surface model. We adopt the effective surface Hamiltonian

\begin{equation}
    \label{eq:surf_ham}
    \hat{h}_{0} = A(k_{y}\hat{\sigma}_{x} - k_{x}\hat{\sigma}_y) + \bm{M}\cdot\bm{\hat{\sigma}}
\end{equation}

in the \(\hat{\sigma}_{z}\) basis, where \(A  = \hbar v_{f}\) and \(\hat{\sigma}_{x,y,z}\) are the Pauli matrices acting on the spin states of the Hamiltonian. In order to model domain walls, the in-plane component of the magnetisation is allowed to rotate around the \(\hat{z}\) axis, i.e. Bloch domain walls. We also consider a small canting of the magnetisation towards the \(\hat{z}\) direction, such that the resulting exchange field is given as \(\bm{M} = (M_{x}(y), M_{y}(y), M_{z})\). This exchange field has exactly the same effect as the Zeeman energy.

Owing to the spatially varying exchange field, \(k_{y}\) is no longer a good quantum number. We make the substitution \(k_{y} \rightarrow -i\partial_{y}\) to account for this spatial variation and to solve for the confined surface states of this model. In doing so, we obtain two differential equations, one for each spin state,

\begin{equation}
    \label{eq:surf_spin}
    \begin{split}
    (E^{2} - M_{z}^{2})\psi_{\sigma}(y) = [&(-iA\partial_{y} + M_{x}) + i\sigma(Ak_{x} - M_{y})]\\&[(-iA\partial_{y} + M_{x}) - i\sigma(Ak_{x} - M_{y})]\psi_{\sigma}(y)
    \end{split}
\end{equation}

\noindent where \(\sigma=\pm 1\) corresponds to the two \(\hat{\sigma}_{z}\) basis states at energy \(E\). In the case of an infinitely wide strip, one can use the Jackiw-Rebbi argument \cite{jackiw_solitons_1976} to show that there must be a surface state at \(E = \pm M_{z}\) satisfying the 1D Dirac equation

\begin{equation}
    [(-iA\partial_{y} + M_{x}) - i\sigma(Ak_{x} - M_{y})]\psi_{\sigma}(y) = 0
\end{equation}

\noindent which admits solutions of the form

\begin{equation}
    \label{eq:infstrip_soln}
    \psi_{\sigma}(y) \propto \exp(-i\int_{\mathbb{R}} \tilde{M}_{x}\text{d}y -\sigma\int_{\mathbb{R}}(k_{x} - \tilde{M}_{y})\text{d}y)
\end{equation}

\noindent where \(\tilde{M}_{x,y} = M_{x,y}/A\). Depending on the exact form of \(M_{y}\) only one of \(\psi_{\uparrow/\downarrow}\) will satisfy the boundary condition \(\lim_{y\rightarrow\infty} \psi_{\sigma}(y) = 0\). However, as previously mentioned, our main interest is in solutions for finite width systems. In the following we will demonstrate the existence of low energy flat-bands within the framework of QW physics, in agreement with the topological arguments for their existence made in the introduction. We will adopt the terminology used in \cite{lu_dirac_2020} and describe the QW formed at the DW as a Dirac QW. In the first instance, we will consider a sharp domain wall of the form \(M_{y} = M(2\theta(y) - 1)\), where \(\theta(y)\) is the Heaviside step function and \(M\) is a constant. We choose \(M>0\) and note that the \(M>0\) and \(M<0\) cases are related by a unitary transformation satisfying \(h_{0}(k_{x},-i\partial_{y},M) = \sigma_{y}R_{y}h_{0}(k_{x},-i\partial_{y},-M)R_{y}^{\dagger}\sigma_{y}\), where \(R_{y}\) is a reflection operator that takes \(y \rightarrow -y\) such that \(\partial_{y} \rightarrow -\partial_{y}\) and \(M_{y} \rightarrow -M_{y}\). We expand equation (\ref{eq:surf_spin}) 

\begin{equation}
    \label{eq:domain_wall_QW}
    \begin{split}
        (E^{2} - M_{z}^{2})\psi_{\sigma}(y) &= [-A^{2}\partial_{y}^{2} + (Ak_{x} - M_{y})^2 + \sigma\partial_{y}M_{y}]\psi_{\sigma} \\
        &= [-A^{2}\partial_{y}^{2} + V(y)]\psi_{\sigma}(y)
    \end{split}
\end{equation}

to obtain the 1D time-independent Schr{\"o}dinger equation with the spin-dependent effective potential \(V(y) = (Ak_{x} - \text{sgn}(y)M)^2 + 2\sigma M\delta(y)\) where \(\text{sgn}(y)\) gives the sign of \(y\). In the case of an infinitely wide strip, solving this equation recovers the solution given in (\ref{eq:infstrip_soln}). A finite strip of length \(L\) can be modelled by adding an infinite potential in the region \(\abs{y} > L/2\). In this case, we must satisfy the boundary conditions \(\psi_{\sigma}(y=\pm L/2) = 0\) and continuity of the wavefunction at \(y=0\). The lowest energy bound states are found at \(E=\pm M_{z}\) and are given by,

\begin{equation}
    \label{eq:finite_strip_soln}
    \psi_{\sigma}(y) = 
    \begin{cases}
        \frac{\psi(0)\sinh{\left(\lambda_{+}(y+\frac{L}{2})\right)}}{\sinh{\left(\lambda_{+}\frac{L}{2}\right)}} & y < 0 \\
        \frac{\psi(0)\sinh{\left(\lambda_{-}(y-\frac{L}{2})\right)}}{\sinh{\left(\lambda_{-}\frac{L}{2}\right)}} & y \geq 0
    \end{cases}
\end{equation}

\noindent where \(\lambda_{\pm} = \abs{k_{x} \pm \tilde{M}}\). In the case of the infinite strip only one of the two spin states satisfies the boundary conditions, the finite strip now hosts spin polarised flat-bands at \(E=\pm M_{z}\) for \(\abs{k_{x}} < \tilde{M}\), with the states comprising the bands localised along the length of the DW (that is, along the line \(y=0\)) in the sample.

Of course, while a sharp DW can prove instructive, such a DW is unlikely to be energetically stable and the magnetisation vector will switch over a greater range. We now consider the case where the DW varies smoothly over a long distance. In the case of the finite strip we can model a smooth DW, to first approximation, as

\begin{equation}
    (M_{x},M_{y}) = 
    \begin{cases}
        (0,-M), & y < -l, \\
        M\left((1-(\frac{y}{l})^{2})^{\frac{1}{2}}, \frac{y}{l}\right), & -l \leq y \leq l, \\
        (0,M), & y > l,
    \end{cases}
\end{equation}

While linear variation of the in-plane exchange field may appear crude, it can be seen as the lowest order term in a realistic domain wall such as \(M_{y} = M\tanh{y/l}\) and therefore valid in the limit of wide DWs. To calculate the energy spectrum of this system, we define the operators \(\hat{a} = (-iA\partial_{y} + M_{x}) + i(Ak_{x} - M_{y})\). The commutation relation of these operators satisfies

\begin{equation}
    \begin{split}
        [\hat{a},\hat{a}^{\dagger}] =  2A\frac{\partial M_{y}}{\partial y}
    \end{split}
\end{equation}

\noindent using the canonical commutation relation and the commutator identity \([f(X),Y] = [X,Y]\frac{\partial f}{\partial X}\). Far from the edges of the system within the region \(\abs{y} < l\), the Hamiltonian is given by

\begin{equation}
    \hat{H} = 
    \begin{pmatrix}
    M_{z} & \sqrt{\frac{2AM}{l}}\hat{c} \\
    \sqrt{\frac{2AM}{l}}\hat{c}^{\dagger} & -M_{z}
    \end{pmatrix},
\end{equation}

\noindent where we have defined the creation and annihilation operators for the harmonic oscillator \(\hat{c} = \sqrt{\frac{l}{2AM}}\hat{a}\) satisfying \([\hat{c},\hat{c}^{\dagger}]=1\), \(\hat{c}\ket{n} = n\ket{n-1}\) and \(\hat{c}^{\dagger}\ket{n} = n\ket{n+1}\) where \(\bra{x}\ket{n}\) are the Hermite polynomials of the quantum harmonic oscillator. Representing \(H\) in the restricted basis \(\{\ket{n-1},\ket{n}\}\) we find \(H_{n}\) equal to

\begin{equation}
    H_{n} = 
    \begin{pmatrix}
    M_{z} & \frac{1}{l_{n}} \\
    \frac{1}{l_{n}}  & -M_{z}
    \end{pmatrix}
\end{equation}

\noindent where \(l_{n} = \sqrt{\frac{l}{2nAM}}\). Solving the secular equation, far from the edges a Landau level (LL) spectrum appears due to the linear variation of the in-plane exchange field with position given by

\begin{equation}
    E_{n} = \pm\sqrt{M_{z}^{2} + \frac{1}{l_{n}^{2}}} .
\end{equation}

Note that, as before there are solutions present at \(E = \sigma M_{z}\) that will now closely resemble a Gaussian wavepacket centred at the domain wall, far from the edges of the sample, \(\psi_{\sigma}(y) \sim \exp(-y^{2}/l^{2})\) for \(\abs{y} < l\). 

\section{\(\bm{k}\cdot\bm{p}\) model for an AFMTI}

Having discussed the formation of bound states in the limits of both sharp and wide DWs, we are now equipped to develop a model AFMTI Hamiltonian with transverse head-to-head DWs. We use a Burkov-Balents type model \cite{burkov_weyl_2011} of a multilayer magnetic TI, in which we consider thin film TI layers separated by wide band gap normal insulator (NI) layers representing the van der Waals gap between TI layers. We may consider the NI to be the vacuum, in which case any states existing within the NI are due to tunnelling between the topological layers either side. Within each TI layer, the non-magnetic part of the Hamiltonian is given as

\begin{equation}
    \label{eq:thin_film}
    \hat{H}_{tf} = \hat{h}_{0}\hat{\tau}_{z} + t_{s}\hat{\tau}_{x}
\end{equation}

in the basis \((\ket{t,\uparrow}, \ket{t,\downarrow}, \ket{b,\uparrow}, \ket{b,\downarrow})^{T}\), where \(t/b\) represent the top and bottom surfaces, \(\uparrow/\downarrow\) represent the up/down spin states in the \(\hat{\sigma}_{z}\) basis, and \(\hat{\sigma}_{x,y,z}\) and \(\hat{\tau}_{x,y,z}\) are Pauli matrices mixing the spins and the surfaces respectively \cite{lu_massive_2010}. The first term, \(h_{0}\), is the surface Hamiltonian of a TI with its surface normal parallel to the \(\hat{z}\) direction, as given in (\ref{eq:surf_ham}). The second term, \(t_{s} = (m + Bk^{2})\), gives the coupling between the top and bottom surfaces. Within the \(j^{\text{th}}\) layer, the interlayer antiferromagnetic coupling gives the in-plane component of the exchange field as \(\bm{M}_{j,\parallel} = (-1)^j(M_{x}, M_{y})^{T} = (-1)^{j}\bm{M}_{\parallel}\). Therefore, depending on the parity of the number of layers, the magnetisation on opposite surfaces will be aligned (odd layers) or anti-aligned (even layers). We once again consider the formation of magnetic DWs along the length of the sample.

Including the magnetic contribution and the coupling between layers, the full Hamiltonian of the multilayer AFMTI is given by

\begin{equation}
    \label{eq:AFMTI_ham}
    H(k_{x},-i\partial_{y}) = 
    \begin{pmatrix}
    H_{+} & T_{z} & 0 & 0 \\
    T_{z}^{\dagger} & H_{-} & T_{z} & 0 \\
    0 & T_{z}^{\dagger} & H_{+} & T_{z} & \cdots \\
    \vdots & & & \ddots \\
    & & 0 & T_{z}^{\dagger} & H_{\eta}
    \end{pmatrix}
\end{equation}

where \(H_{\pm} = H_{tf} \pm \bm{M}_{\parallel}\cdot\bm{\sigma} + M_{z}\sigma_{z}\) is the intralayer Hamiltonian, including a canting of the exchange field towards the \(z\) axis, and \(T_{z} = t_{d}\tau_{-}\) gives the hopping between nearest neighbour TI layers. We have again made the substitution \(k_{y} \rightarrow -i\partial_{y}\) to account for the spatial variation of \(\bm{M}_{\parallel}\).

We are now in the position to consider the effect of the parity of the number of layers on the electronic dispersion of an AFMTI. In order to do so we will consider the simplest, non-trivial examples of even and odd layer parity samples, that is, two and three layer AFMTIs. This is simply for clarity, as the results we obtain for these systems are also applicable for larger even and odd layer systems.

\subsection{Two Layer System}

Using the notation of (\ref{eq:AFMTI_ham}), the Hamiltonian of a two layer system is given as

\begin{equation}
    \label{eq:two_layer}
    H_{2} = 
    \begin{pmatrix}
    H_{+} & T_{z} \\
    T_{z} & H_{-}
    \end{pmatrix}.
\end{equation}

In systems with an even number of layers the magnetisation vectors will be anti-aligned on either surface, breaking surface inversion symmetry. However, this Hamiltonian possesses a \(C_{2}\) symmetry, \([H_{2},P] = 0\), where \(P\) is given by

\begin{equation}
    P = 
    \begin{pmatrix}
    0 & \sigma_{z}\tau_{x} \\
    \sigma_{z}\tau_{x} & 0
    \end{pmatrix}.
\end{equation}

Exploiting this symmetry, \(H_{2}\) (\ref{eq:two_layer}) can be block diagonalised, using the shared eigenbasis of \(H_{2}\) and \(P\). The two sub-blocks \(H^{\pm}\) are given by

\begin{equation}
    \label{eq:two_subblocks}
    \begin{split}
    &H^{\pm} = \\
    &\begin{pmatrix}
        h_{0} + \bm{M}_{\parallel}\cdot\bm{\sigma} + M_{z}\sigma_{z} & t_{s} \\
        t_{s} & -h_{0} + \bm{M}_{\parallel}\cdot\bm{\sigma} + (M_{z} \pm t_{d})\sigma_{z}
    \end{pmatrix},
    \end{split}
\end{equation}

where we have labelled the sub-blocks according to the corresponding eigenvalues of \(P\), \(\pm 1\). Interestingly, this is identical to the thin film Hamiltonian given in (\ref{eq:thin_film}) with an additional spin-dependent surface inversion asymmetry equal to \(\pm t_{d}\). Comparison of (\ref{eq:two_subblocks}) to the surface Hamiltonian, \(h_{0}\), implies that we should expect the appearance of flat-bands at \(E\approx\pm M_{z}\) and \(E\approx\pm(M_{z} \pm t_{d})\).

\begin{figure}[ht!]
    \centering
    \begin{tikzpicture}
        \node[anchor=south west,inner sep=0] (image) at (0,0)
    {\includegraphics[width=8cm]{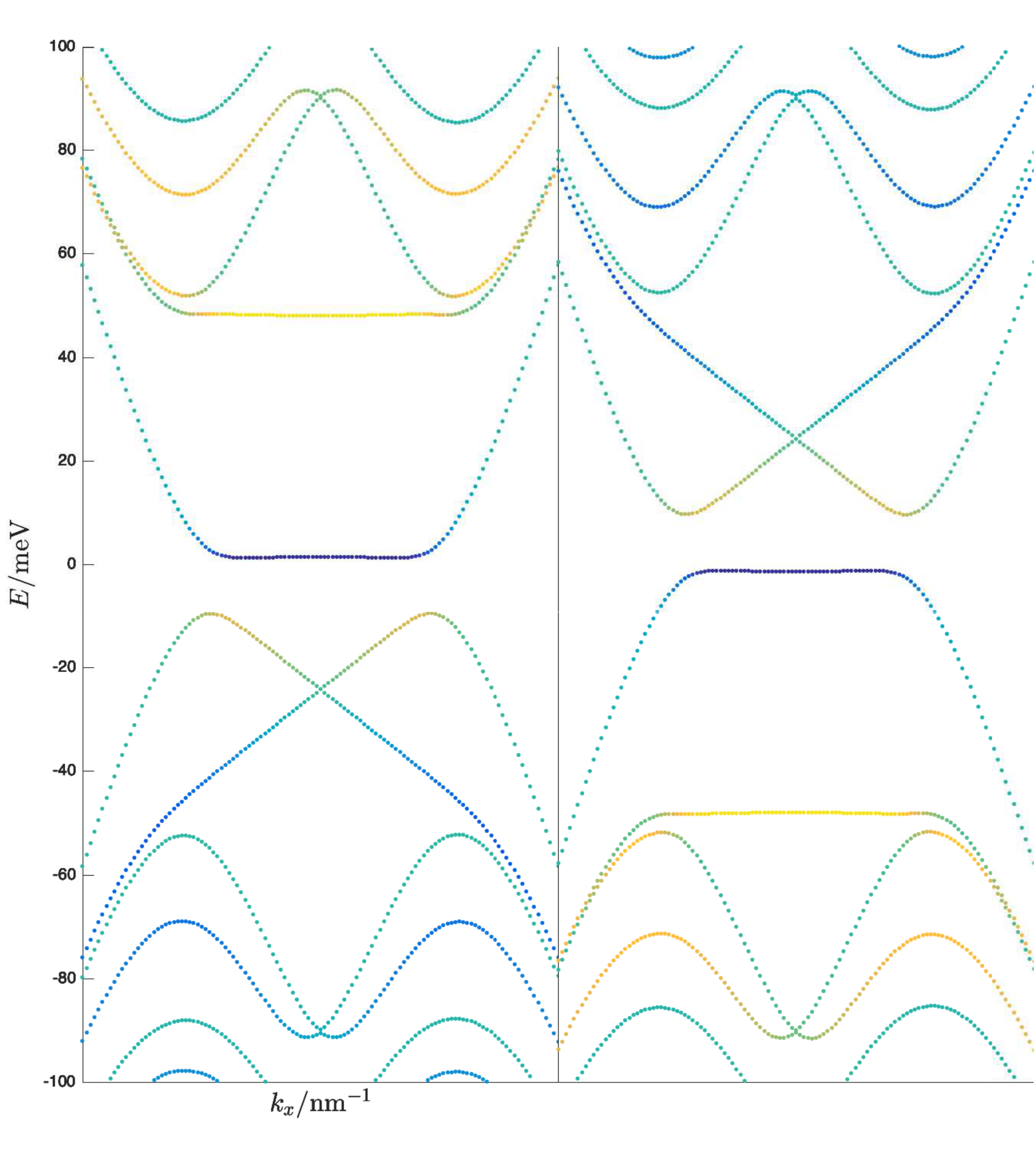}};
        \begin{scope}[x={(image.south east)},y={(image.north west)}]
            %\draw[help lines,xstep=.1,ystep=.1] (0,0) grid (1,1);
            %\foreach \x in {0,1,...,9} { \node [anchor=north] at %(\x/10,0) {0.\x}; }
            %\foreach \y in {0,1,...,9} { \node [anchor=east] at %(0,\y/10) {0.\y}; }
        \end{scope}
    \end{tikzpicture}
    \caption{Electronic dispersion relations from the sub-block Hamilonians \(H^{+}\) (left) and \(H^{-}\) (right). States are coloured according to their spin-z expectation value, \(S_{z} = \bra{\Psi}\sigma_{z}\ket{\Psi}\), with spin up and down given by yellow and blue, respectively.}
    \label{fig:nz=2_H_plus/minus}
\end{figure}

\noindent
However, we should also expect the appearance of edge-states (that is, states localised at \(y = \pm L/2\)) when solving \(H^{\pm}(k_{x},-i\partial_{y})\Psi(y) = E\Psi(y)\) under the boundary conditions \(\Psi(y = \pm L/2) = 0\), based on the solutions obtained from the thin film TI Hamiltonian (\ref{eq:thin_film}). The lack of particle-hole or chiral symmetry in either sub-block, \(H^{\pm}\), means that we shouldn't expect these edge-states around \(E=0\). In order to solve for these edge-states, one can adopt a perturbation theory approach (see appendix) or a numerical scheme. Here we discretise on a square lattice using the finite-difference approximation with a central difference scheme. Unless otherwise stated, all results are based on samples of length \SI{50}{\nano\meter} in the \(\hat{y}\) dimension. Accordingly, parameters in the Hamiltonian are chosen as \(A = \SI{300}{\milli\electronvolt\nano\meter}\), \(m = \SI{25}{\milli\electronvolt}\), \(B = \SI{-100}{\milli\electronvolt\nano\meter^{2}}\), \(t_{d} = \SI{50}{\milli\electronvolt}\) \cite{lu_massive_2010,zhang_crossover_2010}.

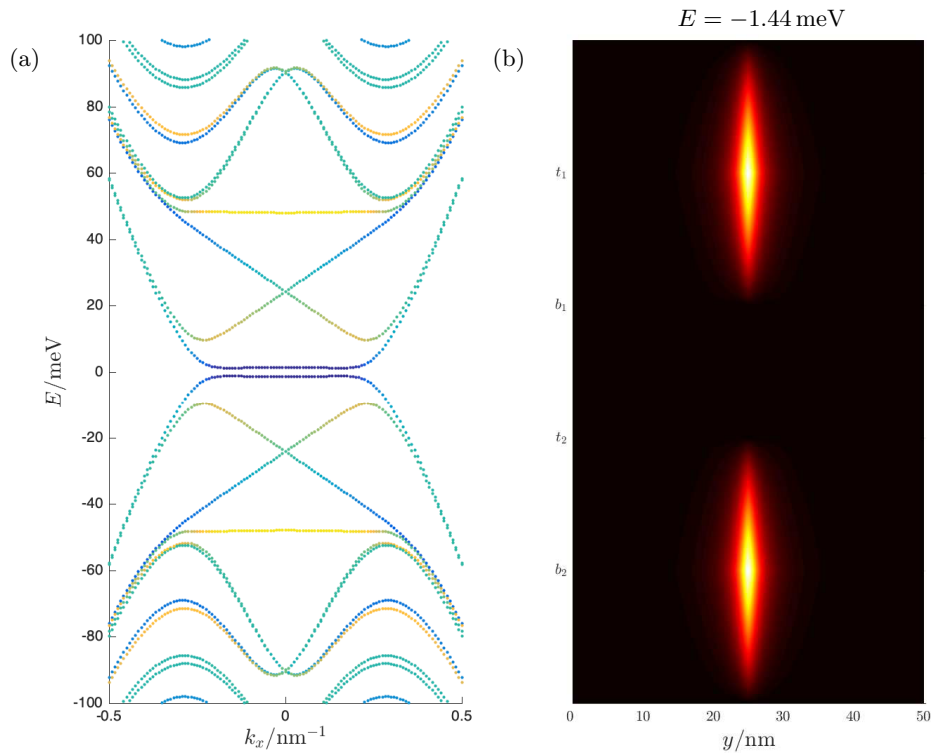
\begin{figure*}[ht!]
    \centering
    \begin{tikzpicture}
        \node[anchor=south west,inner sep=0] (image) at (0,0)
    {\includegraphics[width=0.78\textwidth]{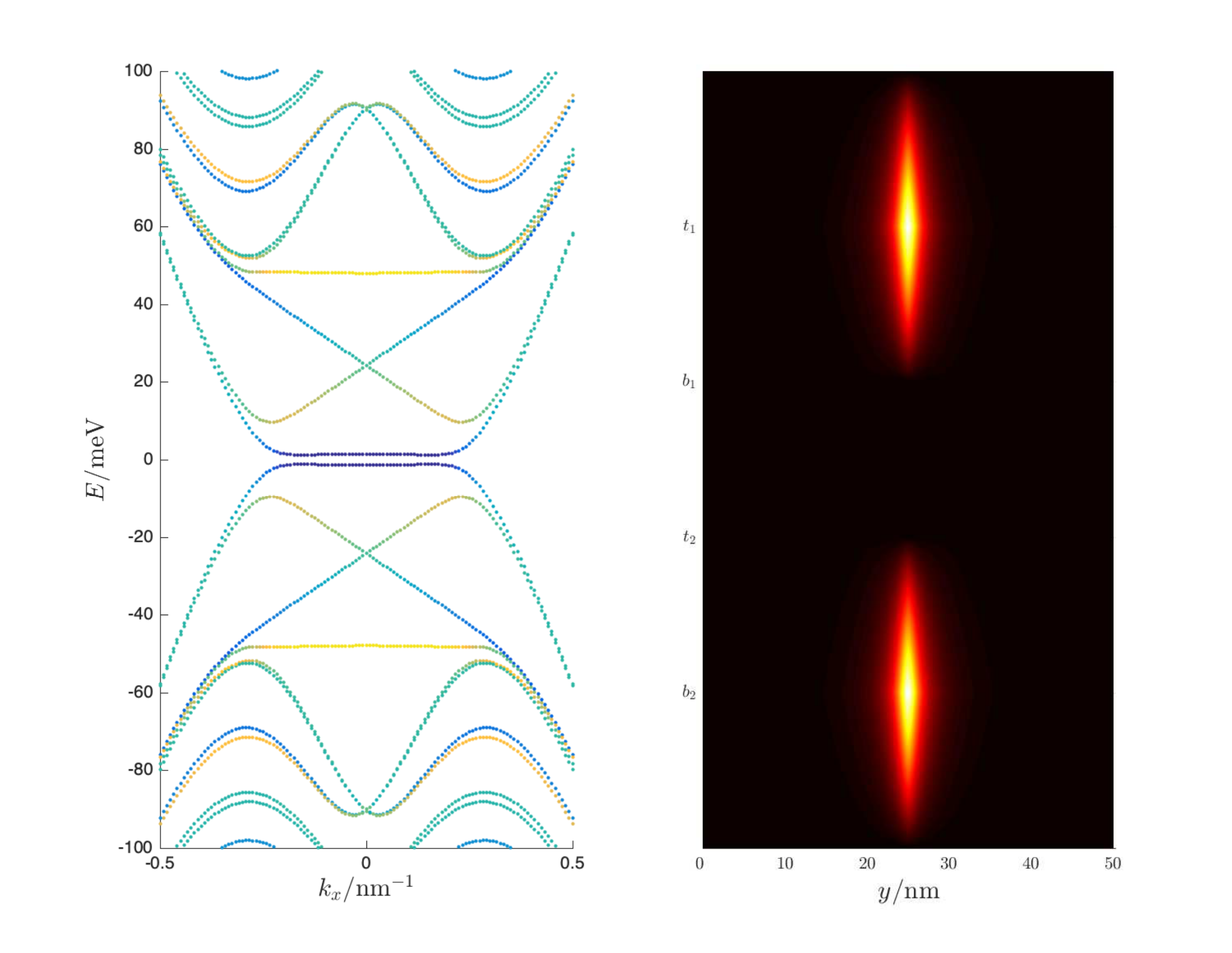}};
    \begin{scope}[x={(image.south east)},y={(image.north west)}]
            %\draw[help lines,xstep=.1,ystep=.1] (0,0) grid (1,1);
            %\foreach \x in {0,1,...,9} { \node [anchor=north] at %(\x/10,0) {0.\x}; }
            %\foreach \y in {0,1,...,9} { \node [anchor=east] at %(0,\y/10) {0.\y}; }
            \draw[black] (0.05,0.9) node {(a)};
            \draw[black] (0.51,0.9) node {(b)};
            \draw[black] (0.75,0.95) node {\(E = \SI{-1.44}{\milli\electronvolt}\)};
        \end{scope}
    \end{tikzpicture}
    \caption{(a) The electronic dispersion of a \SI{50}{\nano\meter} wide, two layer system with \(M_{z} = 0\) and (b) the LDOS around the valence flat-band showing states localised across the domain wall. The vertical position in the multilayer structure is shown along the \(z\) axis where, for example, \(t_{1}\) denotes the top surface of the \(1^\text{st}\) layer. The LDOS was calculated using equation (\ref{eq:LDOS_defn}) with \(k_{B}T = \SI{1}{\milli\electronvolt}\).}
    \label{fig:nz=2_Mz=0_ldos}
\end{figure*}

\begin{figure}
    \centering
    \includegraphics[width=8cm]{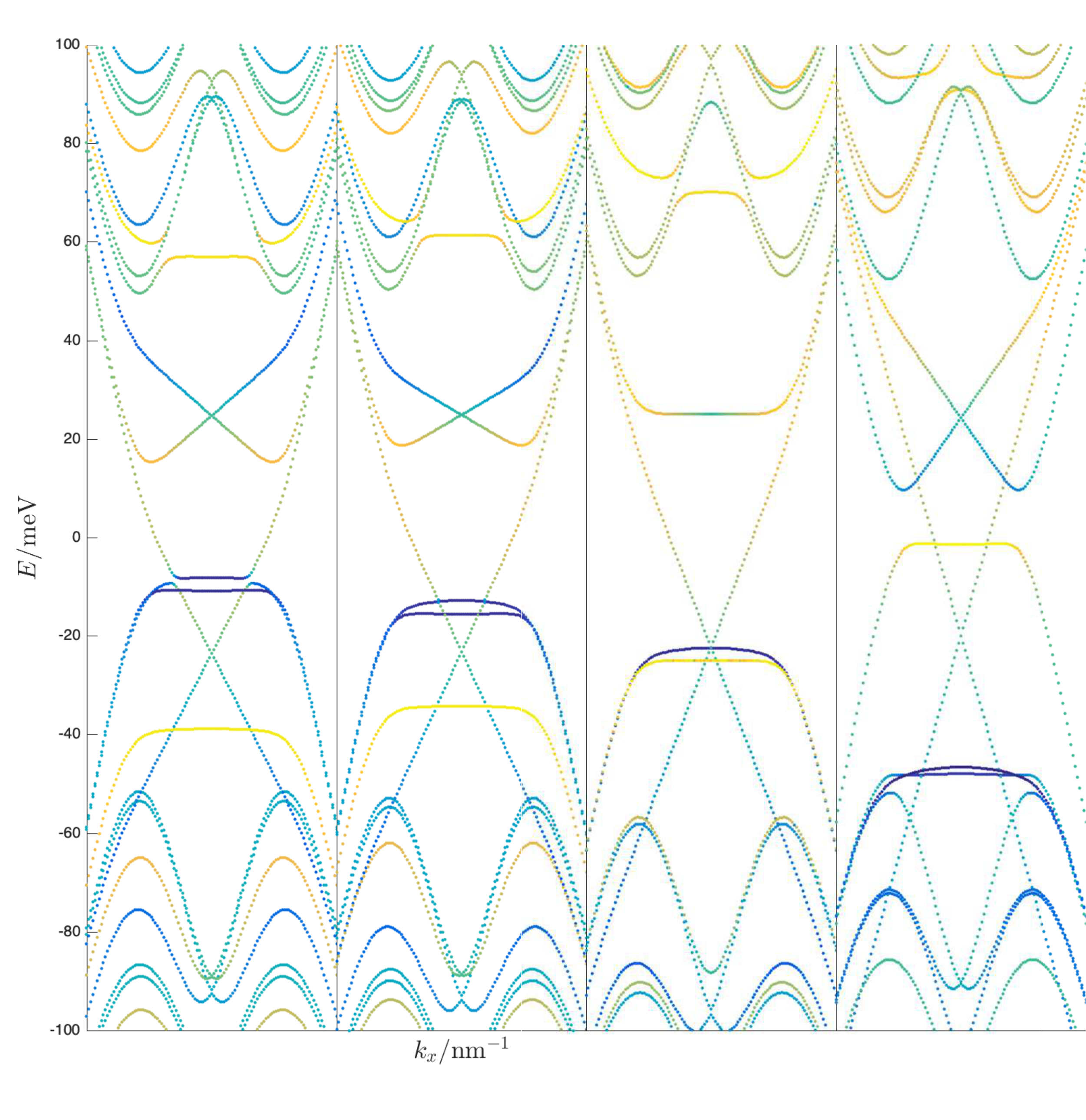}
    \caption{The effect of \(M_{z}\) on a two layer system. From left to right, the electronic dispersion relations shown are \(M_{z} = 10,15,25,50\SI{}{\milli\electronvolt}\).}
    \label{fig:nz=2_effect_Mz}
\end{figure}

Figure \ref{fig:nz=2_H_plus/minus} shows the energy dispersions of the separate sub-blocks, \(H^{\pm}\), with a sharp domain wall, \(M_{y} = 2M(\theta(y) - 1)\), and \(M_{z} = 0\). We focus on the case \(M > 0\) as it is related to the \(M < 0\) case by a unitary transformation. The effect of this transformation is simply to flip the spin-polarisation of the flat-bands. Spin-polarised flat-bands appear in both band structures around \(E = 0\) and \( E = \pm t_{d}\). As discussed above, linearly dispersing bands of surface states are formed away from \(E = 0\). These bands are formed from the remnants of flat-bands that would have otherwise formed at \(E = 0\) and \(E = \pm t_{d}\), had it not been for the intralayer coupling term, \(t_{s}\). As such, the Dirac point of the surface state bands in each sub-block is located around \(E \approx \pm t_{d}/2\). The strength of the coupling between these edge-states and the proximate flat-bands can be controlled by tuning the energy difference between them, using \(M_{z}\), or using the intralayer coupling term, \(t_{s}\). The full band structure of the two layer system is shown in Figure \ref{fig:nz=2_Mz=0_ldos} along with the local density of states (LDOS) around the valence flat-band, that is, the highest occupied flat-band below \(E = 0\). The LDOS is calculated as

\begin{equation}
    \label{eq:LDOS_defn}
    \rho(y,z,E) = \sum_{n}\eta(E - E_{n})\abs{u_{n}(k_{x},y,z)}^{2}
\end{equation}

\noindent
where \(E_{n}\) and \(u_{n}(k_{x},y,z)\) are the energy and eigenstate of the \(n^{\text{th}}\) band, respectively, and \(\eta(E) = -\frac{\partial}{\partial E}f(E)\) is a thermal broadening function, where \(f(E)\) is the Fermi-Dirac distribution. As the complete spectrum of \(H_{2}\) is composed of the \(H^{+}\) and \(H^{-}\) spectra, there is only a small direct band gap between the two low energy flat-bands of each block. By varying the coupling between the top and bottom surfaces of each TI layers using the parameter \(m\), which appears as the wavevector independent term in the intralayer coupling \(t_{s}\), the band gap can be enhanced by controlling the interaction between the flat-bands and the edge-states. The band gap can be increased to terahertz scale energies, making this an ideal platform to study strong correlations within flat-bands by non-destructive means. Turning our attention to the LDOS in Figure \ref{fig:nz=2_Mz=0_ldos}(b), as demonstrated using the surface model (\ref{eq:surf_ham}) the flat-band states are strongly localised around the domain wall. In this case, the \(C_{2}\) rotation symmetry of the two layer system leads to a large density of states on the top and bottom surfaces of the sample. A small electric field applied along the growth direction can break this symmetry, biasing one side of the sample with respect to the other.

\begin{figure*}[ht!]
    \centering
    \begin{tikzpicture}
        \node[anchor=south west,inner sep=0] (image) at (0,0)
    {\includegraphics[width=0.78\textwidth]{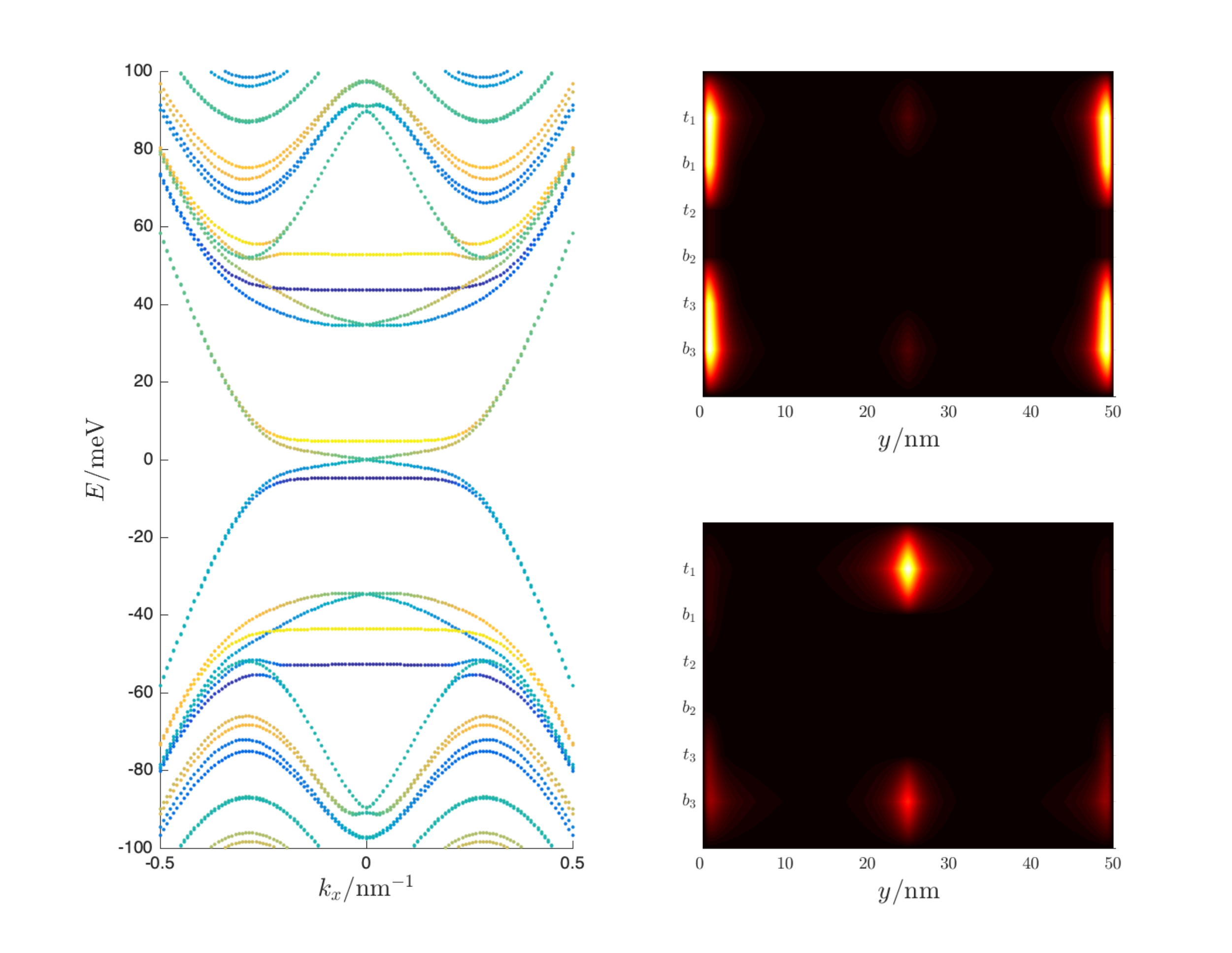}};
    \begin{scope}[x={(image.south east)},y={(image.north west)}]
            %\draw[help lines,xstep=.1,ystep=.1] (0,0) grid (1,1);
            %\foreach \x in {0,1,...,9} { \node [anchor=north] at %(\x/10,0) {0.\x}; }
            %\foreach \y in {0,1,...,9} { \node [anchor=east] at %(0,\y/10) {0.\y}; }
            
            \draw[black] (0.05,0.9) node {(a)};
            \draw[black] (0.51,0.9) node {(b)};
            \draw[black] (0.75,0.95) node {\(E = 6.73\times10^{-5}\SI{}{\milli\electronvolt}\)};
            \draw[black] (0.51,0.43) node {(c)};
            \draw[black] (0.75,0.48) node {\(E = \SI{-4.75}{\milli\electronvolt}\)};
        \end{scope}
    \end{tikzpicture}
    \caption{(a) The electronic dispersion of a \SI{50}{\nano\meter} wide three layer sample with \(M_{z} = \SI{5}{\milli\electronvolt}\) and (b) the LDOS around the energies of the low energy edge-states and (c) the valence flat-band showing states localised across the domain wall. The LDOS was calculated using equation (\ref{eq:LDOS_defn}) with \(k_{B}T = \SI{1}{\milli\electronvolt}\)}
    \label{fig:nz=3_Mz=0_ldos}
\end{figure*}

\begin{figure}
    \centering
    \includegraphics[width = 8cm]{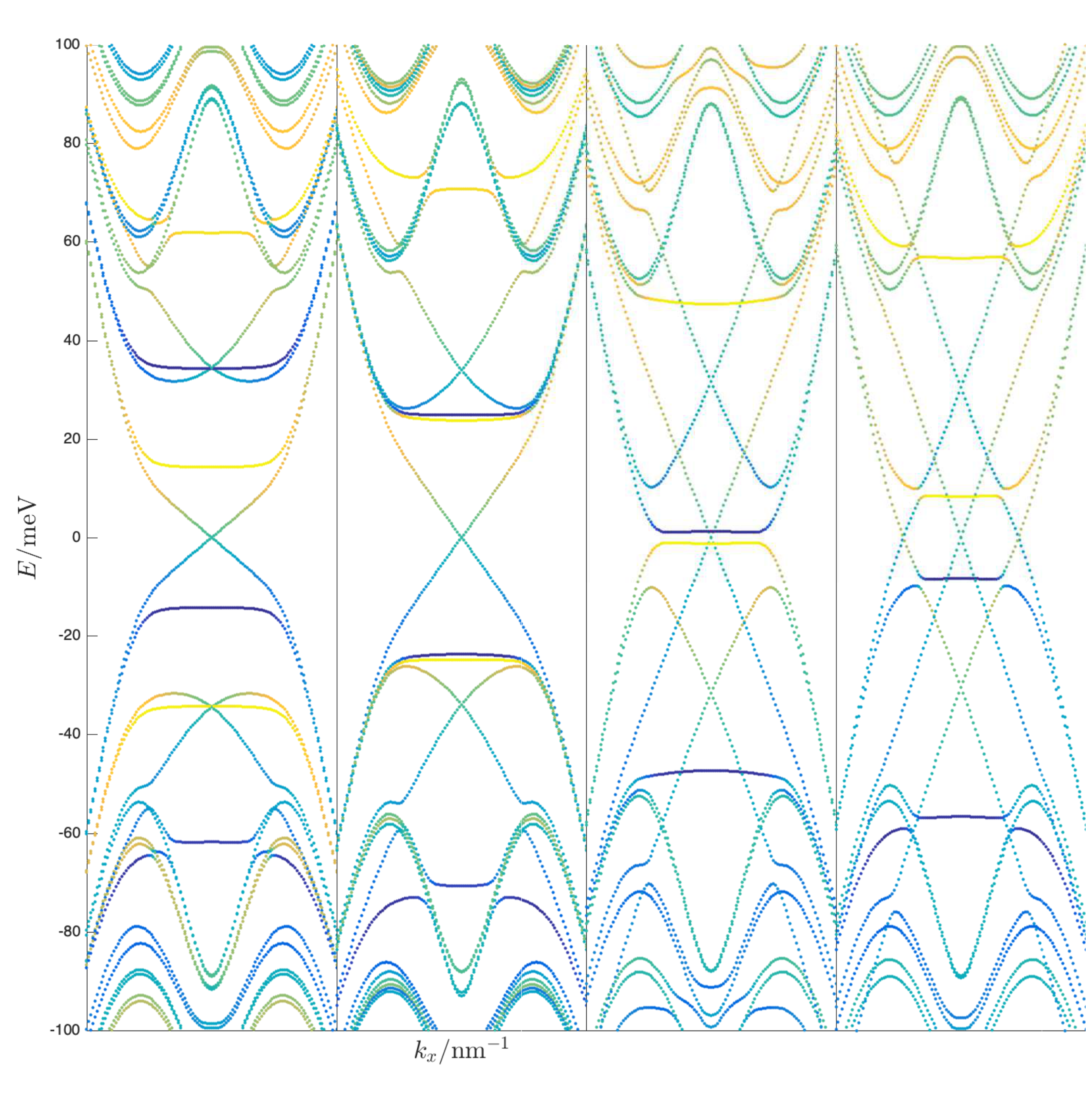}
    \caption{The effect of \(M_{z}\) on a three layer sample. From left to right, the electronic dispersion relations shown are \(M_{z} = 15,25,50,60\SI{}{\milli\electronvolt}\).}
    \label{fig:nz=3_effect_Mz}
\end{figure}

Figure \ref{fig:nz=2_effect_Mz} shows the effect that \(M_{z}\) has on the two layer system. The cases \(M_{z} > 0\) and \(M_{z} < 0\) are related via a unitary transformation and so we present results only on the former. Moderate values of \(M_{z}\) shifts the position of the spin down (up) flat-bands downwards (upwards), i.e. the Zeeman effect, reducing the gap in the \(H^{+}\) block between the low energy flat-band and the edge-states. Increasing \(M_{z} > t_{d}/2\) removes the anti-crossing between the flat-band and the edge state at \(E \approx -t_{d}/2\) in the \(H^{+}\) sub-block, leading to a pair of linear dispersing edge-states. However, the behaviour of the edge-states in the \(H^{-}\) block is strikingly different. The remnants of the flat-bands comprising these bands shift in energy, leading to a reduction in the group velocity until the edge-states are themselves within a set of flat-bands at \(M_{z} = t_{d}/2\). A further increase in \(M_{z}\) leads to a complete inversion of the \(H^{-}\) edge state bands and, eventually, the same behaviour as observed in the \(H^{-}\) sub-block, that is, a removal of the anti-crossing between the edge-states and the spin-polarised flat-band and the formation of a linearly dispersing edge state centred at \(E \approx t_{d}/2\).

\subsection{Three Layer System}

Again, in the notation of (\ref{eq:AFMTI_ham}), the Hamiltonian of a three layer system is

\begin{equation}
    H_{3} = 
    \begin{pmatrix}
    H_{+} & T_{z} & 0 \\
    T_{z}^{\dagger} & H_{-} & T_{z} \\
    0 & T_{z}^{\dagger} & H_{+}
    \end{pmatrix}
\end{equation}

Here, as in all odd layer systems, the magnetisation vector aligns on the top and bottom surfaces. We will exploit this inversion symmetry in demonstrating the particle-hole symmetry of the system, which satisfies \(\Gamma H_{3}(k_{x},-i\partial_{y}) \Gamma^{\dagger} = -H_{3}(-k_{x},-i\partial_{y})\) where \(\Gamma\) is the particle-hole operator defined as

\begin{equation}
    \Gamma = 
    \begin{pmatrix}
    0 & 0 & \sigma_{y}\tau_{y} \\
    0 & \sigma_{y}\tau_{y} & 0 \\
    \sigma_{y}\tau_{y} & 0 & 0
    \end{pmatrix}
    K
\end{equation}

and \(K\) is the complex conjugation operator. In this case, \(\Gamma^{2} = 1\). The presence of particle-hole symmetry within this model ensures that surface states will exist around \(E = 0\), that is, low-energy, topologically non-trivial states. 

The electronic dispersion of a three layer system is shown in Figure \ref{fig:nz=3_Mz=0_ldos}(a) with \(M_{z} = \SI{5}{\milli\electronvolt}\) to clearly show the formation of low energy edge-states between the spin-polarised flat-bands at \(E \approx \pm M_{z}\). Also shown are the LDOS of low energy edge-states, Figure \ref{fig:nz=3_Mz=0_ldos}(b), and the valence flat-band at \(E \approx -M_{z}\), Figure \ref{fig:nz=3_Mz=0_ldos}(c). In contrast to the flat-band states in the two layer model, the low energy flat-band states here are strongly localised around the DW on either the top or bottom surface due to the reflection symmetry present within the three layer sample. Due to thermal broadening the LDOS in Figure \ref{fig:nz=3_Mz=0_ldos}(c) shows a small intensity on the bottom surface, however as the temperature is decreased the LDOS will become more highly localised on the top surface only. Higher energy flat-bands correspond to DW states localised within bulk layers, away from the top and bottom surfaces. Hereafter, these states will be referred to as bulk flat-bands. It is also notable that the low energy edge-states are only localised across the top and bottom layers of the three layer sample. Two more pairs of crossed bands exist away from \(E = 0\) (giving three edge-states in total, corresponding to the number of layers in the sample), which host edge-states localised across the bulk layers of the sample. Therefore, we shall also refer to these states as bulk layer edge-states.

Analogously to the behaviour of the two layer model, we can tune the coupling between these bulk edge-states and the corresponding flat-bands by varying the parameter \(M_{z}\). Again, we only consider the case \(M_{z} > 0\) and present our results in Figure \ref{fig:nz=3_effect_Mz}. Similarly to the two layer system, the effect of moderate values of \(M_{z}\) is a Zeeman effect. As the low energy flat-bands are shifted away from \(E = 0\), the group velocity of the low energy edge-states increases. At larger values of \(M_{z}\) the bulk flat-bands are migrate towards \(E = 0\) and beyond, eventually removing the anti-crossing between these flat-bands and the bulk edge-states. Therefore, at large values of \(M_{z}\), the resultant band structure is three uncoupled Dirac cones bounded by a series of flat-bands.

While it is possible in the two layer model to induce a band gap on the order of \SI{}{\milli\electronvolt} by tuning the parameters in the Hamiltonian, this is not the case for a three layer sample due to the low energy topological states. However, we may exploit the finite size effect \cite{zhou_finite_2008} in order to gap these topological states. 

\section{Conclusion}

In summary, we have discussed the behaviour of spin-polarised flat-bands due to head-to-head transverse DWs in an AFMTI with ground-state planar magnetism. The limiting behaviours of sharp and slowly varying domain walls have been discussed, using a surface Hamiltonian, and we have shown that the functional form of the bound states localised at DWs changes from exponential decay to a Gaussian. The main finding of this work is that the parity of the number of layers of a multilayer AFMTI strongly affects the electronic dispersion. In particular odd layer systems possess particle-hole symmetry, resulting in the presence of low energy topologically non-trivial states. By contrast, a band gap exists between two low energy flat-bands in even layer systems where there are no symmetry protected, low energy topological states. This gap is dependent on the material dependent intralayer coupling term in the AFMTI Hamiltonian and is well within the range of terahertz (THz) radiation.

When many-body interactions are accounted for, we expect that the narrow energy range of these flat bands will put this material system within the strongly-correlated regime. This may lead to emergent novel topological phases of matter, but will also facilitate the study of strong correlations between Dirac fermions. It is interesting to note that the characteristic energy gaps in samples with even numbers of layers naturally lend themselves to being studied in detail by THz radiation in a non-destructive manner. This, along with the existence of multiple growth technologies capable of synthesising thin films of this material system to a high degree of atomic precision, make this a versatile platform to investigate strongly correlated phenomena in topological systems.

\begin{acknowledgments}
NBD acknowledges financial support from the Hitachi Cambridge Laboratory through iCASE studentship RG97400 (voucher 17000178) and the Engineering and Physics Research Council (EPSRC). The authors would also like to thank Dr James Aldous and Dr Peter Newton for valuable input and discussions.
\end{acknowledgments}

\appendix
\section{Edge-states in a Two Layer System}

To solve for the edge-states in a two layer model we set \(k_{x} = 0\) and consider a sharp head-to-head domain wall in the Hamiltonian given by (\ref{eq:two_subblocks}). Following this, we act upon \(H^{\pm}\) with the unitary transformation given by

\begin{equation}
    U = e^{i\tau_{y}\frac{\pi}{4}}
    \begin{pmatrix}
    \mathbb{1} & 0 \\
    0 & \sigma_{y}
    \end{pmatrix}
\end{equation}

to give

\begin{equation}
    UH^{\pm}U^{\dagger} = H^{\pm}_{0} + H^{\pm}_{1}
\end{equation}

where \(H^{\pm}_{0}\) is treated as the zeroth order Hamiltonian and \(H^{\pm}_{1}\) is the perturbation, given by

\begin{equation}
    \begin{split}
        H^{\pm}_{0} &= Ak_{y}\mu_{x} + M_{y}\mu_{y} + t_{s}\mu_{y}\eta_{z} \mp \frac{t_{d}}{2}\mu_{z} \\
        H^{\pm}_{1} &= -(M_{z} \pm \frac{t_{d}}{2})\mu_{z}\eta_{x}
    \end{split}
\end{equation}

where \(\mu_{x,y,z}\) and \(\eta_{x,y,z}\) are Pauli pseudospin operators. First, solving for the unperturbed Hamiltonian, the eigenstates, \(\psi_{\mu,\eta}\) must satisfy

\begin{equation}
    \label{eq:edge_state_diff}
    \begin{split}
    \left(E^{2} - \frac{t_{d}^{2}}{4}\right)\psi_{\mu,\eta} = [&k_{y} - i\mu M_{y} - i\mu\eta t_{s}] \\&[k_{y} + i\mu M_{y} + i\mu\eta t_{s}]\psi_{\mu,\eta}
    \end{split}
\end{equation} 

where \(\mu,\eta = \pm 1\). By expanding the right-hand side of (\ref{eq:edge_state_diff}) and using the trial wavefunction \(\psi_{\mu,\eta} = e^{\lambda y}\), we find the roots \(\pm\lambda_{\mu,\eta,\pm}\) at \(E = \mp\mu t_{d}/2\) given by

\begin{equation}
    \lambda_{\mu,\eta,\pm} = \frac{1}{2B}\left(- A \pm \sqrt{A^{2} + 4B(m + \eta M_{y})}\right)
\end{equation}

Note that we must make the substitution \(M_{y} = -M\) for \(y < 0\) and \(M_{y} = M\) for \(y > 0\). Therefore, the solution for \(\psi_{\mu,\eta}\) is a linear combination each solution to (\ref{eq:edge_state_diff}) in each half space. The coefficients may be determined by fulfilling the relevant boundary conditions at \(y = \pm L/2\) and \(y = 0\), however we will not do that here. The resulting solutions are a pair of edge-states at \(E = \mp t_{d}/2\) and two states localised around the DW at \(y = 0\) at \(E = \pm t_{d}/2\). Introducing the perturbation Hamiltonian, \(H_{1}^{\pm}\), the interfacial states are shifted in energy by an amount \(\sim \pm (M_{z} \pm t_{d}/2)\) owing to their strong overlap. By contrast, the overlap between edge-states is extremely small in wide samples since states are localised on opposite edges. As a result, edge-states remain located around \(E \approx \mp t_{d}/2\). 

\bibliography{references}

\end{document}